\documentclass[letterpaper,twocolumn]{esapub} 

\usepackage{epsfig}
\newcommand{\be}{\begin{equation}}
\newcommand{\ee}{\end{equation}}
\newcommand{\bea}{\begin{eqnarray}}
\newcommand{\eea}{\end{eqnarray}}
\newcommand{\Bea}{\begin{eqnarray*}}
\newcommand{\Eea}{\end{eqnarray*}}
\newcommand{\pa}{\partial}

\newcommand{\na}{\nabla}
\title{Radiative hydrodynamics in
the highly super adiabatic layer of stellar evolution
models}
\author{F.J.Robinson}
\author {P.Demarque}
\author {S.Sofia}
\affil{Yale University, New Haven CT USA}
\author{K.L.Chan}
\affil{Hong Kong University of Science \& Technology, Hong Kong, China}
\author{Y.-C.Kim}
\affil{Yonsei University, Seoul, South Korea}
\author{D. B.Guenther}
\affil{St Mary's University, Halifax, N.S., Canada}

\begin{document}
\maketitle

\begin{abstract}
We present results of  three dimensional simulations of the uppermost part of the sun,
at 3 stages of its evolution.
Each  model includes
physically realistic radiative-hydrodynamics (the Eddington approximation
is used in the optically thin region), varying  opacities and a realistic
equation of state (full treatment of the ionization of H and He).
In each evolution model, we investigate
a domain, which starts at the top of the photosphere and ends just inside the convection zone
(about 2400 km in the sun model). This
includes all of the super-adiabatic
layer (SAL). 
Due to the different
positions of the three models in the $log (g) $ vs $log T_{eff}$ plane,
the more evolved models have lower density atmospheres.
The reduction in density causes the amount of overshoot into the radiation layer, to be
greater in the more evolved models.
\end{abstract}

\section{Introduction}
In most of the convection zone, heat transport can be 
modelled quite well by using the mixing length theory. Due to vigorous turbulent 
mixing, the entropy 
is almost constant. This well mixed layer has 
an extremely small super-adiabatic temperature gradient ($\na - \na_{\rm ad}  \approx 10 ^{-8}$) 
and is practically opaque (the optical depth is greater than $10^4$).  In such a layer,  radiative transport 
accounts for a very small fraction of the heat flow.
This changes further up.  
Near the top of the convection zone the gas density is very low, so the enthaply 
flux is not big enough to carry all of the outward flowing heat.  Radiation must 
carry most of the heat flux. 
To enable radiation to transport this heat flux,  
the temperature gradient must increase significantly. 
This region, in which $\na -\na_{ad}$ is of order unity, 
is known as the super adiabatic layer (SAL). 

The SAL is important for the following reasons:

\begin{itemize}
\item
It is the site of maximum amplitude of the p-mode oscillations (particularly the high frequency modes).
\item
As most of the entropy change occurs within this layer, 
it plays an important role in  determining the radius of stellar models (Larson 1974).
(Note $dS/dlnP=c_{\rm p} [\na - \na_{\rm ad}]$.)
\item
The amount of overshoot into the enveloping photosphere 
depends on the  structure of the SAL.
\end{itemize}

\begin{figure}
\centering
\epsfig{file=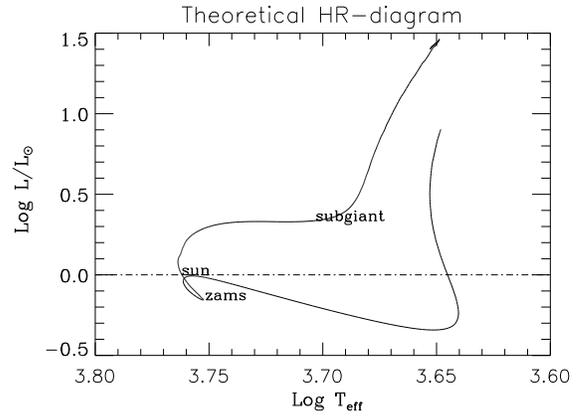, width=8cm}
\caption{Evolutionary track for the sun from the pre-main sequence stage to the giant branch}
\label{hr}
\end{figure}

In this poster we will describe three simulations of stellar atmospheres each centered around the SAL. We 
consider the sun at three 
different stages of its evolution, 
namely the
ZAMS, present sun and subgiant.
These are shown as three points in the theoretical HR diagram, figure \ref{hr}.

\section{Mathematical model}
\label{ns}
\subsection{Hydrodynamic equations}
In the outer layers of the sun the Mach number,
$(v/v_{\rm s})^2$ can be  of order  unity (Cox and Giuli 1968) ($v$ is the flow velocity 
and $v_{\rm s}$ is the sound speed).
In such an environment, the governing  hydrodynamic equations are the fully compressible
Navier Stokes equations (cf, Kim et al 1995). These are modified to include 
radiative energy transport by the insertion of the  $Q_{\rm rad}$ term (described later) into the energy equation.
The full set of governing equations are:
\bea 
\pa \rho / \pa t &=& - \na \cdot  {\bf \rho v }\nonumber\\
{\bf  \pa \rho v} / \pa t & =& - {\bf\na \cdot \rho  v v}  
- \na  P
+ {\bf \na \cdot \mbox{\boldmath$ \Sigma$ }}
+ \rho {\bf g } \nonumber\\
\pa E /  \pa t& =&   - \na \cdot [(E+P) {\bf v 
-  v \cdot \mbox{\boldmath$ \Sigma$ }} 
+ f ]\nonumber\\\nonumber
&& + {\bf \rho v \cdot g} + Q_{\rm rad}\nonumber
\eea 
where $ E = e + \rho {v^2} /2$ is the total energy density and $\rho, {\bf v }, P, e$ 
and ${\bf g}$, are the density,  velocity, pressure, specific internal energy
and acceleration due to gravity, respectively.  
Ignoring the coefficient of bulk viscosity (Becker 1968), the viscous stress tensor
for a Newtonian fluid
is $\Sigma_{ij}=\mu(\pa v_i/\pa x_j+\pa v_j/\pa x_i)-2\mu/3(\na \cdot {\bf v})\delta_{ij}$.
In LES,
the
dynamic viscosity $\mu$  is increased so that it represents the effects of Reynolds
stresses on the unresolved or sub-grid scales (Smagorinsky 1963),
\bea
\mu=\rho(c_\mu\Delta)^2(2 \mbox{\boldmath $\sigma : \sigma$})^{1/2}.
\nonumber
\eea
The colon inside the brackets denotes tensor contraction of the rate of strain tensor
$\sigma_{ij} = (\na_i v_j+\na_j v_i)/2$.
The SGS eddy coefficient $c_\mu$, is set to 0.2, the value for incompressible
turbulence, and $\Delta$ is an estimate of the local mesh size.
The  present formulation ensures that the grid Reynolds number $\Delta \times v/\nu$  is of order
unity everywhere.
To handle shocks, $\mu$ is
multiplied by $1 +C \cdot (\na \cdot {\bf v})^2$, where
the constant $C$  is made as small as possible to maintain numerical stability.
As $\mu$ is dependent on the horizontal divergence, any 
large horizontal velocity gradients are smoothed out by the increased viscosity.
The diffusive flux ${\bf f} = -(\mu / {\rm Pr}) T \na S $, where the horizontal mean of $\na S \leq 0$ 
i.e. the convection zone, 
and ${\bf f} = -(\mu c_{\rm p}/{\rm Pr}) \na T $ where  the horizontal mean of $\na S \geq  0$ 
i.e. the radiation zone.  The Prandtl number Pr $=\nu / \kappa $ ($\nu$ is the kinematic viscosity and 
$\kappa$ is the thermal diffusivity) is  1/3. Due the inclusion of radiative energy transport,
the effective Pr is actually much smaller 
and not constant.

\subsection{Radiative energy transport}
In the deeper part of the domain, radiative transfer is treated by the diffusion approximation, 
\bea
Q_{\rm rad} = \na \cdot \left[\frac{4acT^3}{3\kappa\rho}\na T \right], \nonumber
\eea
where $\kappa$ is the Rosseland mean opacity, 
$a$ is the Boltzmann constant and $c$ is the speed of light.

In the shallow region,  
the photon mean free path is at least one tenth 
of the depth of the atmosphere 
so the diffusion approximation may not apply. Instead
$Q_{\rm rad}$ is computed as  
\bea
Q_{\rm rad} = 4 \kappa \rho (J - B)\nonumber 
\eea
where mean intensity $J$ is computed by using the generalised three-dimensional Eddington 
approximation (Unno and Spiegel 1966), 
\bea 
\na \cdot \left( \frac{1}{3\kappa\rho} \na J \right) - \kappa  \rho J + \kappa \rho B = 0. \nonumber
\eea 
This is exact for isotropic radiation,
and in non-equilibrium the Eddington approximation describes the optically 
thick and thin 
regions exactly. 

\section{Numerical methods}
The simulation domain is a small box of aspect ratio 1.5, which includes
the photosphere and top part of convection zone. In the sun model
this represents $6.936 \times 10^{10} {\rm cm} < R < 6.960 \times 10^{10} {\rm cm}$.
Each of the three models span about 8 pressure scale heights.

The governing equations are discretised in cartesian coordinates on  $80 ^ 3$ uniformly 
spaced grid points.
Using a code developed by Chan and Wolff (1982), an implicit scheme (the Alternating
Direction Implicit Method on a Staggered grid or ADISM) relaxes the fluid to a self consistent
thermal equilibrium.  
In the  fully relaxed layer, the energy flux leaving the
top of the box is within 5 $\%$ of the input flux at the base.
Next a second order explicit method (Adams Bashforth time
integration) gathers the statistics of the time averaged state. The statistical integration
time is about 2500 seconds of solar surface convection, and requires about a  million time steps. 
On an 667 MHz Alpha processor, each integration step requires about 10 seconds of CPU time.
Consequently each simulation takes at least 3 months to run.  

\subsection{Modelling a stellar atmosphere}
A Standard Solar model, calculated with the Yale Stellar Evolution model (Guenther \& Demarque 1997),  is used  to
compute the initial stratification i.e. run of pressure, temperature, density, internal energy, for the box. 
The model  uses  realistic physics,   Alexander low temperature opacities
and the OPAL opacities and Equation of State (Hydrogen and helium ionisation zones are included).
The hydrodynamical  simulations use identical opacities and  e.o.s,  as used in the 1-d stellar model
(Kim \& Chan 1998).
The horizontal boundaries are  periodic, while  
the 
vertical boundaries are stress free.
A constant heat flux flows through the base, and the top is a perfect conductor.               
To ensure  mass,  momentum and energy are fully conserved, 
we use impenetrable (closed) 
top and bottom boundaries.

\begin{figure}
\centering
\epsfig{file=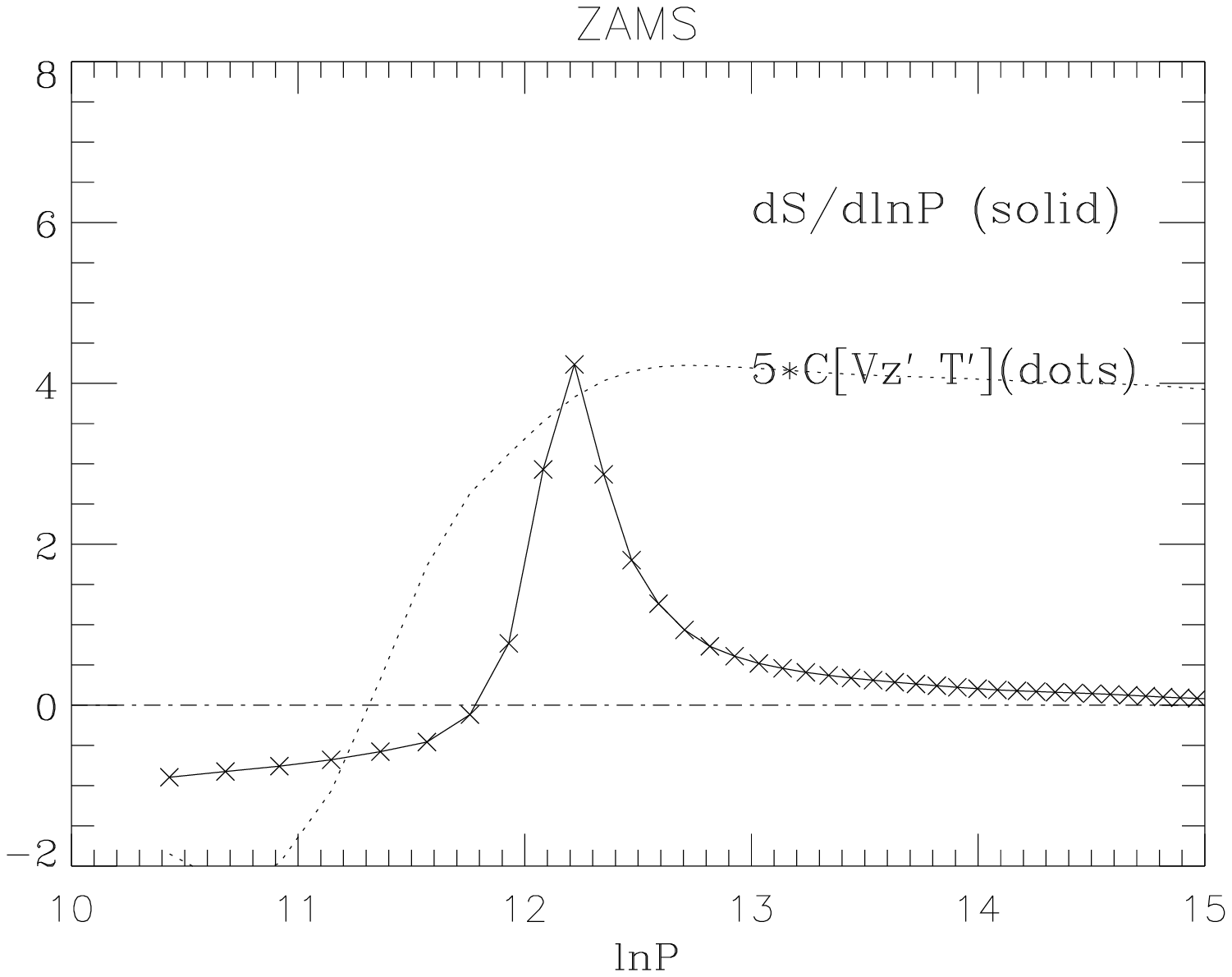, width=8cm}
\label{fig2}
\caption{Horizontally averaged entropy gradient and correlation between vertical velocity and temperature    
vs depth, for the ZAMS model. 
The crosses represent the actual grid points.
The normalised correlation is rescaled to fill the same figure.
}
\epsfig{file=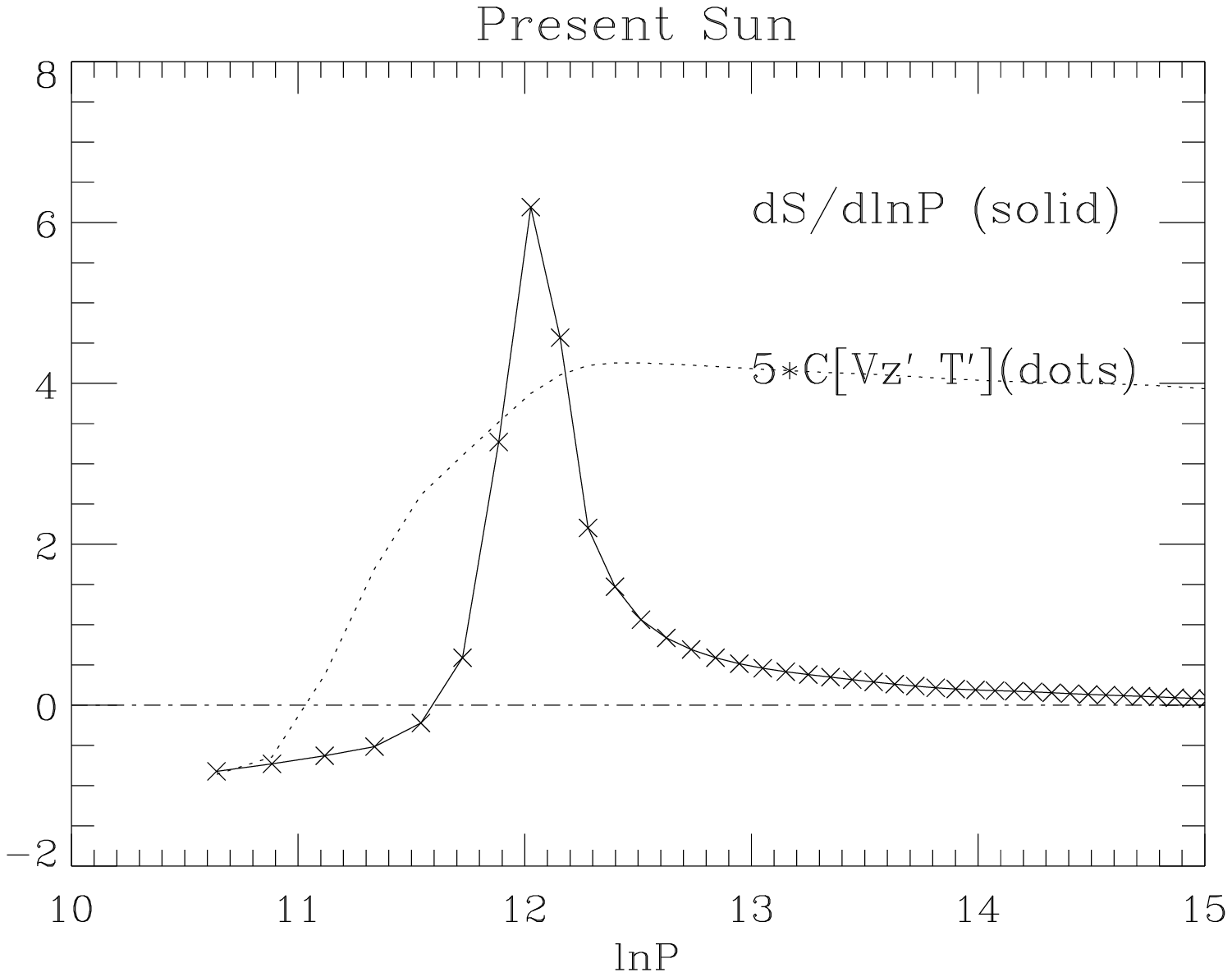, width=8cm}
\label{fig3}
\caption{Same as figure 2 for the sun model.}
\epsfig{file=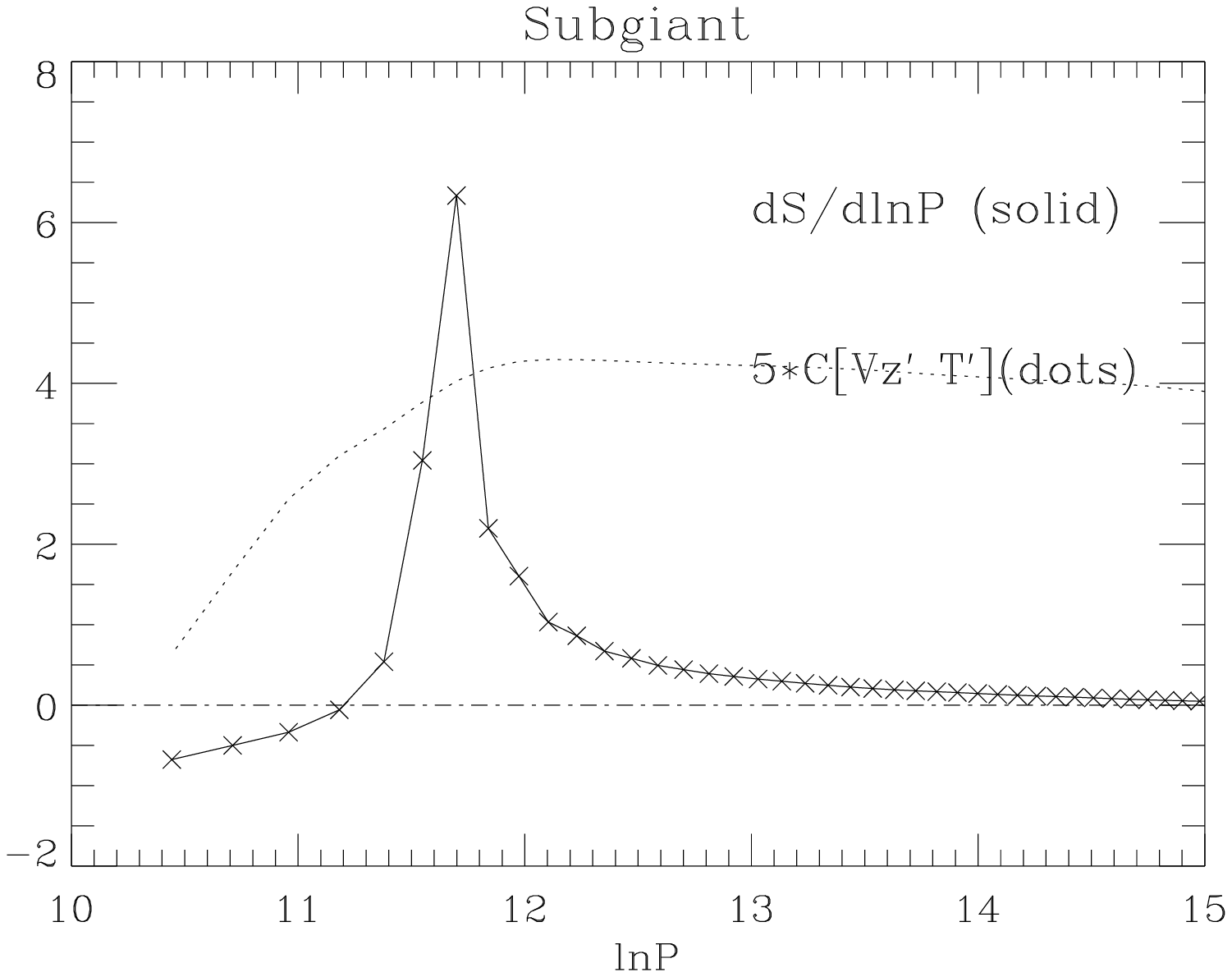, width=8cm}
\label{fig4}
\caption{Same as figure 2 for the subgiant model.}
\end{figure}
\begin{figure}
\centering
\epsfig{file=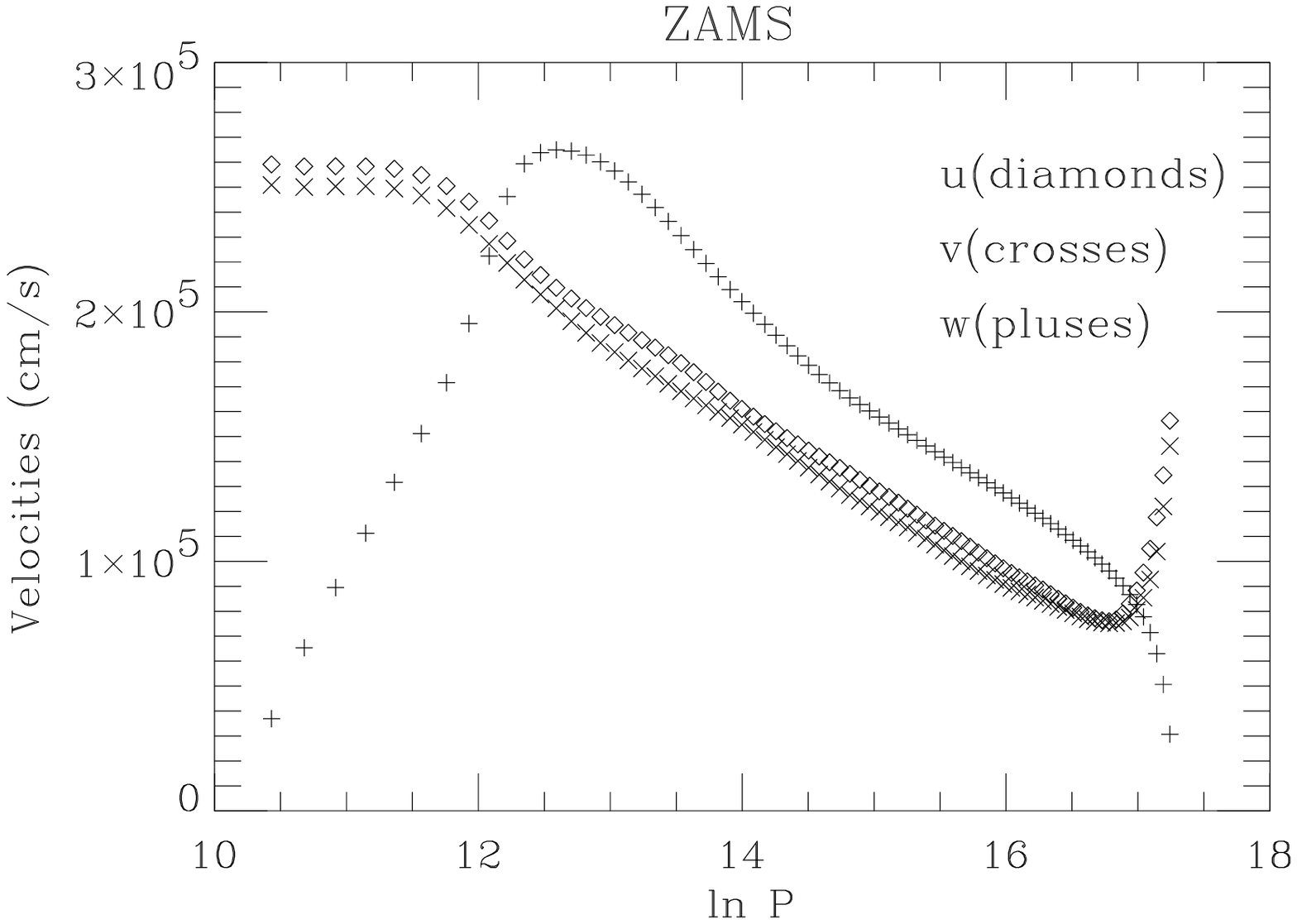, width=8cm}
\label{fig5}
\caption{For  the ZAMS model, the turbulent velocities in the 
horizontal ($u$ and $v$) and vertical ($w$) directions, versus depth.
The closeness of the two horizontal velocities confirms that the 
simulation is statistically converged.}
\epsfig{file=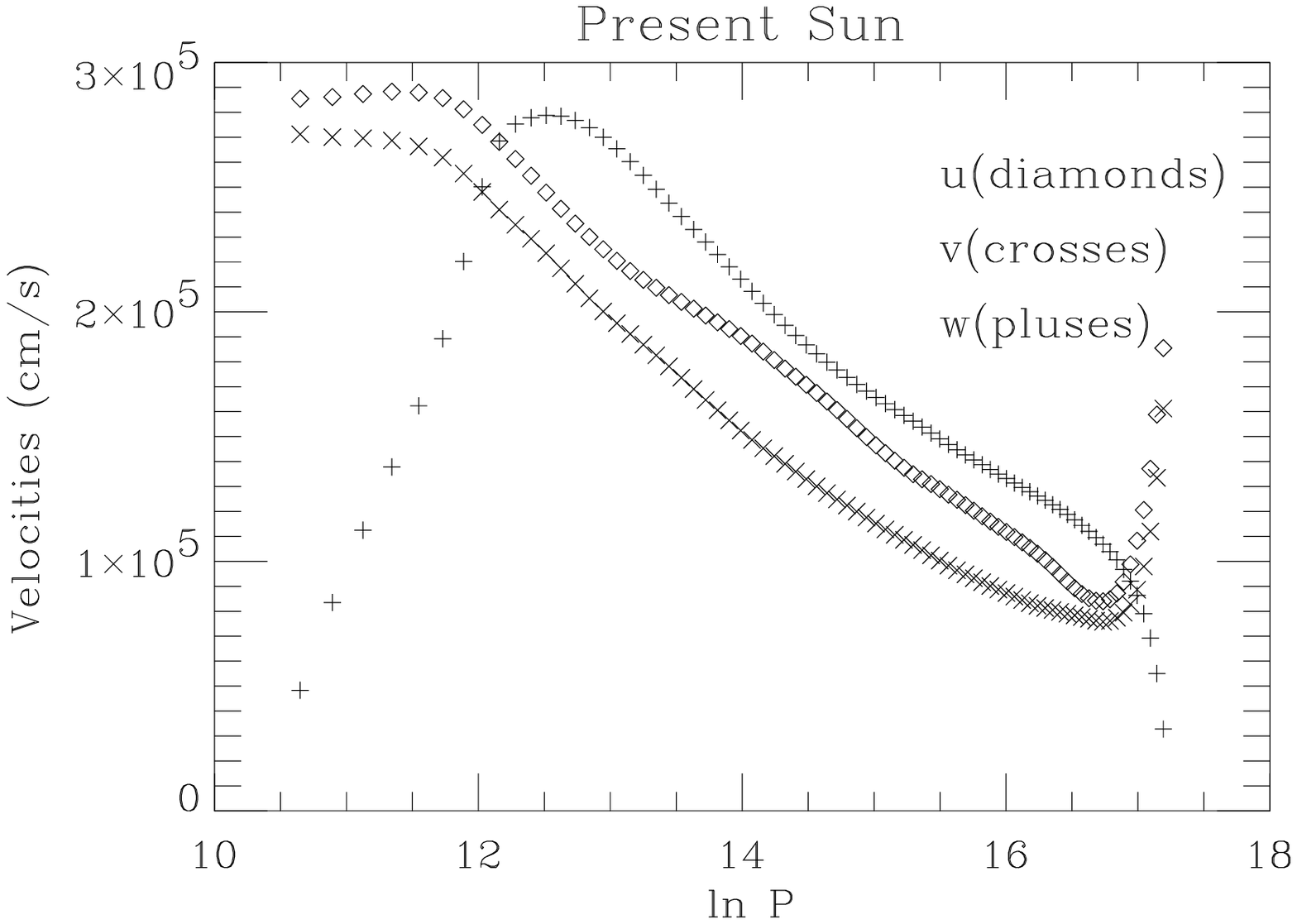, width=8cm}
\label{fig6}
\caption{Same as figure 5 for the sun model.}
\epsfig{file=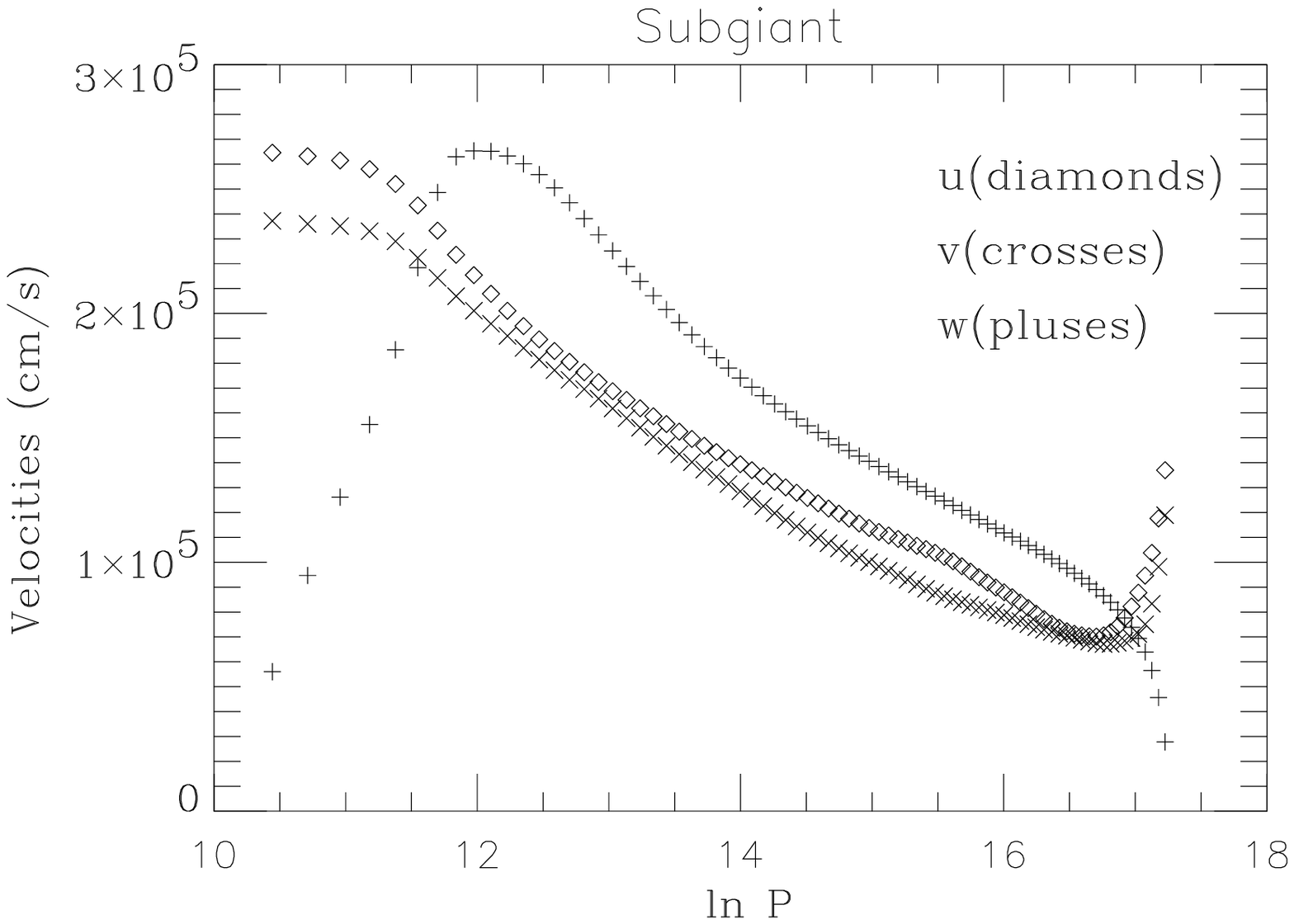, width=8cm}
\label{fig7}
\caption{Same as figure 5 for the subgiant model.}
\end{figure}

\section{Results and Inferences}
\subsection{Turbulent quantities}
\label{turb}
Each parameter, which consists of a mean and a fluctuating part, is computed
from the 3-d statistically averaged flow.  For a given parameter $X$, the
turbulent part is approximated by the variance,

$x = \sqrt{ \overline{X^2} - {\overline{X}} ^2 }$,

where
the overbar denotes horizontal and temporal averaging, and $X$ is the total
quantity (mean plus fluctuating).
The autocorrelation between two fluctuating quantities 
$X_1'$ and $X_2'$ is computed as 

$C[X_1' X_2'] = \overline{ (X_1-\overline{X_1})((X_2-\overline{X_2})} /x_1 x_2$. 

\subsection{Structure of the SAL in the three models}
Figures 2, 3 and 4 show the entropy gradient and the autocorrelation  between vertical 
velocity fluctuation  and temperature, $C[V_z' T']$,  vs depth,
using $\ln {\rm P}$ as the unit of depth. 
  
The SAL peak is higher and sharper in the subgiant model compared to the  ZAMS model. 
This is  because density is much lower in the SAL region for the subgiant, 
than it is for the ZAMS. Convection is less efficient, and thus $\na - \na_{\rm ad}$ (or 
similarly $dS/dlnP$) needs to be steeper and higher, so that radiation can carry more heat.
Generally, the SAL regions are radially further outwards in progressively more 
evolved models.

The difference  between the top of the convection
zone, as defined by $dS/dz=0$ (Schwarzschild  criterion),  and  the point where the 
enthapy flux, $\rho c_{\rm p} C[V_z' T'] w t$,
is zero,
is one measure of the amount of overshoot into the photosphere.
The figures (which exclude the two outermost levels) show that the overshoot increases slightly between the 
ZAMS and sun models, while extending  high up into the lower density region
in the subgiant model.  
It is important to note that because the grid points are uniformly spaced, 
the resolution of the SAL is worse in the subgiant than in the ZAMS.
As the SAL occurs higher up in the subgiant 
where the pressure scale height is smaller, there are less grid points 
between $\ln{\rm P}$ equals 11 and 12, than between 12 and 13.  Clearly it is very important 
to have enough grid points to model the SAL in 
more evolved stellar models. 
\subsection{Turbulent velocities}
The turbulent velocities 
$u$, $v$ and $w$, computed as described in section \ref{turb} are shown in figures 5,6 and 7. 
The closeness of the turbulent horizontal velocities ($u$ and $v$) indicates the degree of statistical convergence.
The ZAMS and sub-giant models are fully converged, while the sun is close to convergence. 

\section*{Acknowledgement}
 This research was supported in part by NASA grant
NAG5-8406 to Yale University.  Support from the
Creative Research Initiative Program of The Korean Ministry of
Science and Technology (Y.-C. Kim) and from the National Science and
Engineering Research Council of Canada (D. B. Guenther) are also
gratefully acknowledged.
F.J.Robinson acknowledges helpful comments from H. Ludwig in improving this poster.  

\end{document}